\documentstyle[aps,prb,epsfig]{revtex}

\newcommand{\uprule}{\end{multicols}
\noindent \vrule width3.375in height.2pt depth.2pt
\vrule height.5em depth.2pt \hfill \widetext }
\newcommand{\downrule}{\indent \hfill \vrule depth.5em height0pt
\vrule width3.375in height.2pt depth.2pt
\begin{multicols}{2} \narrowtext}

\newcommand{\be}{\begin{equation}}
\newcommand{\ee}{\end{equation}}  
\newcommand{\ba}{\begin{eqnarray}}
\newcommand{\ea}{\end{eqnarray}}
\newcommand{\Ln}{\Lambda_n}
\newcommand{\Lc}{\Lambda_c}
\newcommand{\dpdx}{ {{d\phi} \over dx}}
\newcommand{\dpdt}{ {{d\phi} \over dt}}
\newcommand{\dedx}{ {{d\eta} \over dx}}
\newcommand{\dedt}{ {{d\eta} \over dt}}

\begin{document}
\draft
\title{Numerical Test of Disk Trial Wave function for Half-Filled Landau Level}
\author {S.-R. Eric Yang}
\address {Department of Physics, Korea University, Seoul 136-701, Korea}
\address {Asia Pacific Center for Theoretical Physics, Seoul, Korea}
\author {Min-Chul~Cha}
\address {Department of Physics, Hanyang University, Ansan 425-791, Korea}
\author {Jung Hoon Han}
\address {Department of Physics, U.C. Berkeley, Berkeley, CA 94720 USA }

\maketitle
\begin{abstract}
The analyticity of the lowest Landau level wave functions
and the relation between filling factor and the total angular
momentum severely limits the possible forms of trial wave functions of
a disk of electrons subject to a strong perpendicular magnetic field.
For $N$, the number of electrons, up to $12$ we have tested these disk
trial wave functions for the half filled Landau level
using Monte Carlo and exact diagonalization methods.
The agreement between the results for the occupation numbers
and ground state energies
obtained from these two methods is excellent.
We have also compared  the profile of the occupation
number near the edge with that obtained from a field-theoretical method.  
The results give qualitatively identical edge profiles. 
Experimental consequences are briefly discussed.
 
\end{abstract}
\maketitle

\section{Introduction}
There has been a great interest in two-dimensional 
electron liquid in a strong magnetic field at half-filling.
Its bulk properties have been investigated and several novel features have 
been revealed\cite{halp,read1,recent,pas,lee,read2}.
Electrons form a new exotic metal in which they move effectively
in zero magnetic field despite the presence of an external field \cite{halp}.  
A dipole-like new particle\cite{read1,recent,pas,lee,read2}
with fermionic statistics has been introduced. 
This new state of matter forms a Fermi liquid with a finite
compressibility \cite{halpste}.

The edge properties of this exotic liquid are of great current interest,
both experimentally\cite{gray}
and theoretically\cite{shy,leewen,frad,yang1,yang2,yang3}.
Recently a two-mode model for the edge \cite {leewen} has been proposed 
based on a theory of the electron liquid at $\nu=1/2$ 
consisting of charged Bose and neutral Fermi liquids \cite{lee}.
In this theory the charge mode propagates while the neutral mode
is dispersion-less.
A theory based on the topology of Abelian fractional quantum Hall
states also leads to similar results,
but in this approach two non-propagating modes of a 
purely topological origin exist\cite{frad}.

A disk of electrons consists of a bulk and an edge,
and it is desirable to treat them in the same theoretical framework.
The edge and bulk properties have been also investigated using a
trial wave function for an  electron disk 
at half-filling\cite{yang1,yang2,yang3}.
The trial wave function was constructed using analyticity of
the lowest Landau level wave functions,
and the fact that the filling factor ($\nu$)
is related to the total angular momentum ($M_z$)
through $\nu=N(N-1)/2M_z$.
These conditions severely limit the possible
forms of the wave function \cite{yang1,yang2}.
It was tested that the wave function overlap of this
state with the exact one is essentially one  for $N=5$ \cite{yang3}.  
The computed edge occupation number showed,
in sharp contrast to incompressible Laughlin states, that a tail 
exists in the occupation number beyond the radius of a 
droplet of uniform filling factor $1/2$ \cite{yang2}.
The properties of zeroes 
of the wave functions have been also investigated \cite{yang2}.  
The proposed trial wave function can be used to understand
the dynamic structure factor, which is a physically measurable quantity.
This was done by constructing trial wave functions for
some of the low energy excitations of a disk \cite{yang3}.
One of them is the center of mass (CM) motion,  
which can be interpreted as an edge mode.
Other excitations are identified and their wave functions are also constructed.
They are not edge modes, but are excitations of the entire droplet.
They are new excitations of a droplet at $\nu=1/2$ which do not 
exist in incompressible states.
The dynamic structure factor shows that the disk can support
several low energy excitations, unlike incompressible quantum Hall states.   

In this paper we perform an extensive numerical test of
the proposed trial wave functions for an  electron disk at half-filling.
We find that the previously proposed trial wave functions \cite{yang1,yang2}
work quite well up to $N=8$.  We find that for bigger values of $N$
almost degenerate states with lower energies exist.  
Using these trial wave functions we have computed several physical
properties by performing Monte Carlo simulations,
and have  tested the obtained results against exact diagonalization
results for up to $N=12$.
The agreement between the results for the occupation numbers
and ground state energies obtained from these two methods is excellent.
We have also compared  the profile of the occupation
number near the edge with that of a field theoretical
approach\cite{leewen,frad}.
They give qualitatively identical edge profiles. 

This paper is organized as follows.
In Sec.~II we explain the composite Fermion wave function,
which is a central concept in our trial wave function approach.
Next, in Sec.~III, we give the physical idea behind the construction
of trial wave functions at the principal filling factors 
using composite Fermion wave functions.
In Sec.~IV we perform Monte Carlo simulations 
using the proposed trial wave functions and compared the extracted
physical properties with those of exact diagonalization method.
A field theoretical calculation of the edge occupation number is carried out
in Sec.~V and the obtained result is compared with those of
the exact diagonalization and Monte Carlo results.
Discussions are given in Sec.~VI.  In this paper length and energy
are measured in units of $\ell$ and $e^2/\epsilon\ell$, where $\ell$
is the magnetic length.

\section{Composite Fermion Wave functions}
In this section we explain composite fermion wave function,
which is a central concept in our trial wave function approach.
This concept was introduced by Jain and Kamila \cite{JK}.
Before we discuss our trial wave function approach  
it is necessary to understand this concept first.
So we give a short summary of their result.

\subsection{Physical Model of a Disk}
Let us first set up our model of a disk.
We consider a system of 2D electrons confined by a 
parabolic potential, $V(r)=m_e\Omega^2r^2/2$.
The advantage of this model is
that some exactly known results can be utilized in constructing
the collective modes.
Such a system will form a uniform
density electron disk for sufficiently large $N$.  
The symmetric gauge single-particle
eigenstates $\phi_{n,m}(z)$ 
are conveniently classified by a Landau level index $n$ and an angular 
momentum index $m=-n,-n+1,...$ ($z=x+iy$). 
In the strong field limit the single particle orbitals in the $n=0$ level 
have energies $\epsilon_m=\hbar\omega_c/2+\gamma(m+1)$,
where $\gamma=m_e\Omega^2 \ell^2$.   
The system is invariant under spatial rotations about
an axis perpendicular to the
2D plane and passing through the center of the dot.
It follows that the total angular momentum $M_z$ is a good quantum number.
Eigenenergies may be expressed as a sum
of interaction and single-particle contributions
\be
E_i(N,M_z)=U_i(N,M_z)+\gamma(N+M_z).
\ee
Here $i$ labels a state within a $M_z$ subspace,
and $U_i(N,M_z)\propto e^2/\epsilon \ell$ is
determined by exactly diagonalizing the electron-electron interaction term
in the Hamiltonian within this subspace.
We will present in this paper a different approach
than some of previous  works 
on fractional quantum Hall states  disk geometry \cite{yang4,other}.

\subsection{Composite Fermion Wave functions}

Recently it has been argued that the main physics of strong electron
correlation in quantum Hall systems is that the electrons carry
two Jastrow factors  
\cite{jain}.
\be
\psi[z] = P_{L^{3}}\prod_{j<k}(z_{j}-z_{k})^{2}
\left | \matrix {
   \phi_{n_{1}m_{1}}(z_{1}) & ..... & \phi_{n_{1}m_{1}}(z_{N}) \cr
                 ..... & ..... & .....  \cr 
   \phi_{n_{N}m_{N}}(z_{1}) & ..... & \phi_{n_{N}m_{N}}(z_{N}) \cr }
   \right |
\ee
where $P_{L^3}$ stands for projection to the lowest Landau level (LLL).
It is useful to rewrite this as
\be  
\psi[z] =P_{L^{3}}
  \left | \matrix {
   \phi_{n_{1}m_{1}}(z_{1})J_{1} & ..... & \phi_{n_{1}m_{1}}(z_{N})J_{N} \cr
                 ..... & ..... & .....  \cr
   \phi_{n_{N}m_{N}}(z_{1})J_{1} & ..... & \phi_{n_{N}m_{N}}(z_{N})J_{N} \cr}
   \right | 
   \label{wavef1}
\ee 
In physical terms, the Jastrow factor $J_i=\prod_{k\neq i}(z_i-z_k)$ implies
that each particle carries
with it a correlation hole with respect to all the other particles. 

The projection $P_{L^3}$ is tantamount to replacing the anti-holomorphic
variable $\bar{z}$ with $2(\partial/\partial z)$\cite{gv}.
It is not straightforward to project this wave function to the LLL.
Jain and Kamila (JK)\cite{JK} have proposed the following wave functions
of interacting fermions in the LLL for a given total angular momentum $M_{z}$:
\be
\Psi^{CF}=\left | \matrix {
   \eta_{n_{1}m_{1}}(z_{1}) & ..... & \eta_{n_{1}m_{1}}(z_{N}) \cr
         ..... & ..... & .....  \cr
   \eta_{n_{N}m_{N}}(z_{1}) & ..... & \eta_{n_{N}m_{N}}(z_{N}) \cr}
   \right |
   \label{wavef2}
\ee
\be
  \eta_{nm}(z_i)=
  e^{-\frac{1}{4}\mid z_i \mid^{2}}z_i^{m+n}\partial_i^{n}J_i
\ee
$\eta_{nm}(z)$ is {\it not} a genuine single-particle state as its 
value depends on the position of all the other particles.
Nonetheless to stress the similarity between
Eq.(\ref{wavef1}) and Eq.(\ref{wavef2})JK regard $n$ and $m$ as
the LL index and the angular momentum of the composite fermions (CFs),
respectively.  The composite fermion 
Landau level (CFLL) index is just a convenient nomenclature
for some abstract mathematical property of $\eta_{nm}(z)$.
{\it It should be noted that CFLL
index $n$ is not a real Landau level index, and although
higher CFLL indices appear in this wave function
it lies entirely in the lowest electron Landau level}.
On the other hand, it can be shown that the total angular 
momentum of $\Psi_{CF}$ is $M_z=N(N-1)+M^*$, where
$M^*=\sum_i m_i$ is the sum over angular momentum of CF levels ($N(N-1)$
stems from $N$ Jastrow factors of $J_i$).

JK choose the ground state as the compact state $[N_0, N_1,..,N_{l-1}]$,
where the number of CFs in the {\it i}-th CFLL is $N_i$.  This state is defined to satisfy:
[1] $N_0\geq N_1\geq N_2,...$; [2] $\sum_{i=0}^{l-1}N_i=N$; 
[3] For given $N_i$, all the CF levels 
$m=-i,-i+1,..,N_i-i-1$ are assumed to be occupied.  
For example, for $N_i=5$ the CF occupation numbers are
given by $(\bullet,\bullet,\bullet,\bullet,\bullet,\circ,\circ,...)$ 
in the $i$-th CFLL [An occupied (unoccupied) CF
state is represented by a full (empty) circle.
The angular momentum of a CF state increases from left to right].  
The ultimate justification of choosing these wave functions comes
from  the comparison with 
exact diagonalization results (See Sec. IV.B.1 Wave function overlaps).

We will drop the universal factor 
$\prod_i e^{-|z_i |^2 /4}$ from now on. 
When we write out the wave function of $[N_0, N_1,..,N_{l-1}]$ it has the form
\be
\left|\begin{array}{cccccccc}
J_1 & z_{1} J_1 & ... & z_{1}^{N_0 -1}J_{1} & \partial_1 J_1 &... &
z_{1}^{N_1 -1}\partial_1 J_{1} & ... \\
J_2 & z_{2} J_2 & ... & z_{2}^{N_0 -1}J_{2} & \partial_2 J_2 &... &
z_{2}^{N_1 -1}\partial_2 J_{2} & ... \\
  & & & ... & & & &     \\
J_N & z_N J_N & ... & z_{N}^{N_0 -1}J_{N} & \partial_N J_N & ... &
z_{N}^{N_1 -1}\partial_N J_{N} & ...
       \end{array}\right|.
\ee
Successively higher powers of $\partial_i$ acting on $J_i$ appear as we
move to the right for a given row. The matrix elements are of the
form $z_i^{n+m}\partial_{i}^{n}J_i$. 
One can easily verify that the $1/3$ Laughlin state \cite{laug} 
is given, in this notation, by $[N_0, N_1,..,N_{l-1}]=[N,0,...,0]$: 
\be
  \left | \matrix 
    { J_{1} & z_{1}J_{1} & ..... & z_{1}^{N-1}J_{1}\cr 
      ......& ..... & ..... & .....  \cr
      J_{N} & z_{N}J_{N} & ..... & z_{N}^{N-1}J_{N}\cr}
    \right |
\ee 
It can also be verified that the wave function for the filled electronic
Landau level at $\nu=1$ is given by
$[N_0, N_1,..,N_{l-1}]=[1,1,..,1]$: 
\be
 \left | \matrix
    { J_{1} & \partial_{1}J_{1} & ..... & \partial_{1}^{N-1}J_{1} \cr
      ..... & ..... & ..... & .....\cr
      J_{N} & \partial_{N}J_{N} & ..... & \partial_{N}^{N-1}J_{N} \cr}
    \right |
\ee
JK do not give  specific prescription  for other general filling factors.

\section{Construction of Trial Wave functions for Principle Filling Factors}
Now that we have explained the composite fermion wave functions 
we can proceed to construct  trial wave functions at other principal filling factors.
\subsection{Uniform density droplet}
\begin{figure}
\hskip -0.4cm
\epsfxsize=16cm \epsfysize=16cm \epsfbox{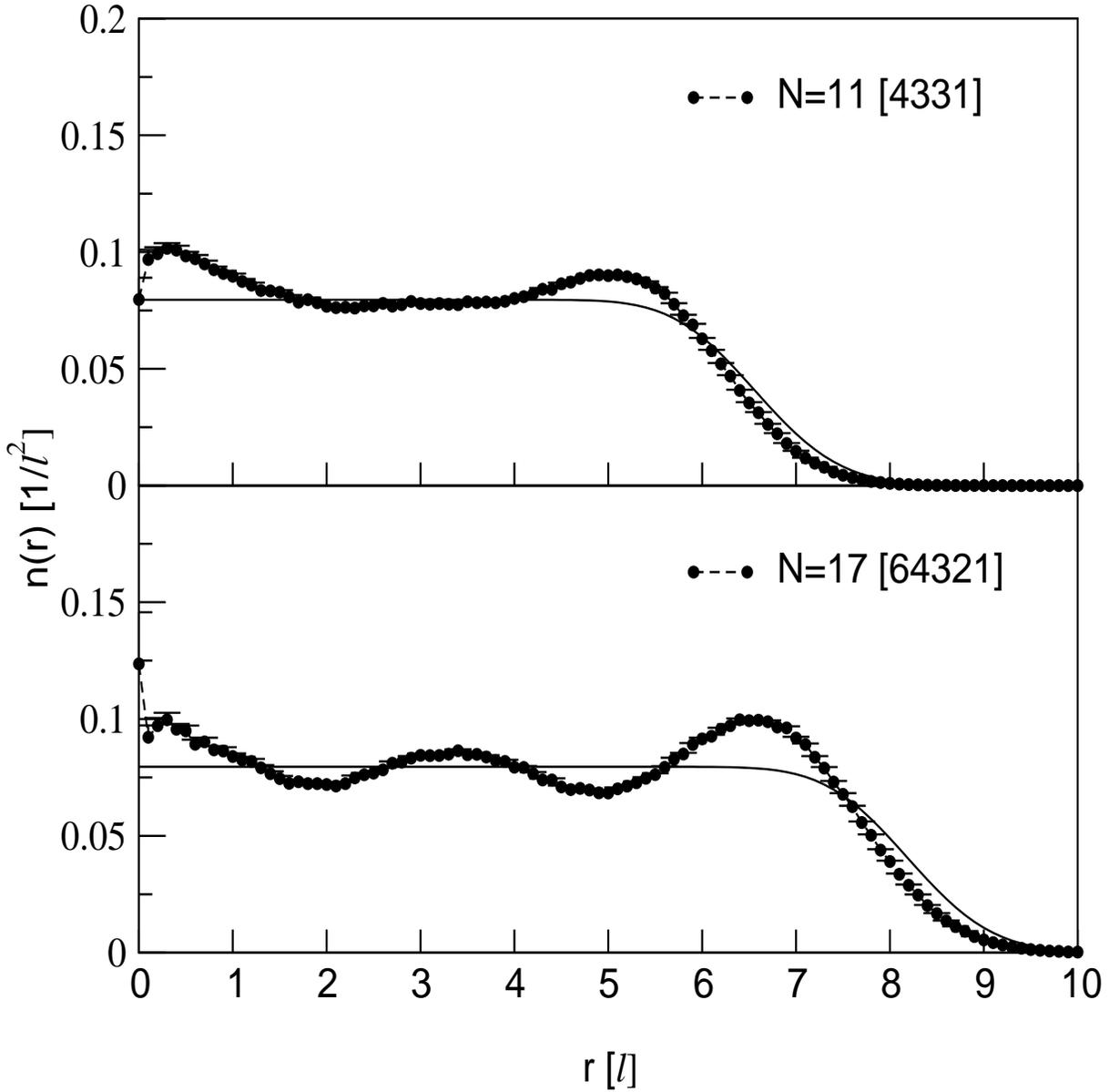}
\caption{The density profiles of droplets at $\nu=1/2$ for $N=11$ and $17$.
Filled circles represent the Monte Carlo calculation of trial wave functions,
while the solid lines represent the expected density profile for the
case that the angular momentum states are evenly occupied ($n_m=1/2$)
up to $m_{ed} = 2N-1$.  Note that, although there are some deviations between the two,
the average value of the Monte Carlo results
agree with the expected value.
Even for the Laughlin states at $(\nu,N)=(1/3,25) (1/5,20) (1/7,15)$ the density
profiles have deviations from the expected value \cite{occ1}.}
\label{den1117}
\end{figure}

The formula 
$\nu=N(N-1)/2M_z$, which relates the filling factor to the
total angular momentum can be shown to be true for Laughlin states
using plasma analogy \cite{laug}.
For other states described by composite fermion wave functions 
numerical results support the validity of this formula.
We have tested numerically the  wave functions $[4,3,3,1]$ and $[6,4,3,2,1,1]$,
which we believe describe
the $\nu=1/2$ state at $N=11$ and $N=17$, respectively.
We have verified for these states 
that the average electron density $n(r)$ inside
dot is approximately  $(1 /2\pi \ell^2)\nu$ (see Fig.~\ref{den1117}).
We have also tested other states $[6,1,...,1]$ and
$[9,1,...,1]$ for $N=11$ and $N=17$ and found similar results.  
We expect from the size dependence that the  agreement should become better 
for $N >>17$ ($17$ is the biggest size we could handle numerically).
Even for the Laughlin states numerical results show that the correct average
value is approached slowly as $N$ increases \cite{occ1}.

\subsection{First method}
Let us first explain our trial wave function approach using 
analyticity of the lowest Landau level wave functions
and the fact that the filling factor is related to
the total angular momentum. 

As a trial wave function for the state at $\nu= 1/2$
Yang and Han\cite{yang1,yang2} have proposed the state 
$[N/2,2,1,...,1]$ ($[(N-1)/2,1,...,1]$) for even (odd) $N$.
It can be  verified that for each value of $N=3,4,5,6,7$
this state is the {\it only} possible compact state satisfying the condition
$\nu=N(N-1)/2M_z$.
The occupation of CFs in these states are listed in Table~\ref{energy}. 
The state $[\frac{N}{2},2,1,....,1]$ has the form 
\be
\Phi^{'}_0= \left | \matrix
    { J_{1} & z_{1}J_{1} & ..... & z_{1}^{\frac{N}{2}-1}J_{1} & 
\partial_{1} J_{1}& ..... & \partial_{1}^{\frac{N}{2}-1}J_{1} \cr
     ..... & ..... & .....& .....  &..... &.....  \cr
      J_{N} & z_{N}J_{N} & ..... & z_{N}^{\frac{N}{2}-1}J_{N} &  
      \partial_{N}J_{N} & ..... & \partial_{N}^{\frac{N}{2}-1}J_{N} \cr}
     \right |
\label{generalwaveftn}
\ee

This state has the following form in terms of
antiholomorphic variable $\bar{z}$
\be
 \left | \matrix
    { J_{1} & z_{1}J_{1} & ..... & z_{1}^{\frac{N}{2}-1}J_{1} &  
\bar{z}_{1} J_{1}& ..... & \bar{z}_{1}^{\frac{N}{2}-1}J_{1} \cr
     ..... & ..... & .....& ..... & ..... &.....  \cr
      J_{N} & z_{N}J_{N} & ..... & z_{N-1}^{\frac{N}{2}-1}J_{N} &  
      \bar{z}_{N}J_{N} & ..... & \bar{z}_{N}^{\frac{N}{2}-1}J_{N} \cr}
     \right |
\ee
Han and Yang\cite{yang2} have also proposed trial wave functions at other
principal filling factors.

\subsection{Second method}
The proposed wave functions can be also constructed using a more physical 
argument based on counting flux numbers and their relation to the total
angular momentum.

Consider the $\nu=1/3$ state $[N]$, which has $3N$ flux quanta.
The angular momentum of this state
is $3N(N-1)/2$.  If two flux quanta are removed from this state 
the number of flux quanta per particle will be $1/\nu'=\frac{3N-2}{N}$.
Then, in the large $N$ limit, the total angular momentum will be given by 
\be
M_z=\frac{1}{\nu'}N(N-1)/2=3N(N-1)/2-(N-1) \approx 3N(N-1)/2-N.
\ee
>From this we see that reducing two flux quanta amounts to reducing
the total angular momentum by $N$.
The resulting state is $[N-1,1]$ since the angular momentum of this state
is $N(N-1)+M^{*}=3N(N-1)/2-N$.
Here we have used the fact that the total angular momentum of composite
fermions $M^{*}$ in the state $[N-1,1]$
is $-1+\sum_{i=0}^{N-2}m_i=-1+(N-1)(N-2)/2$.
If two flux quanta are removed additionally, 
the angular momentum is  again reduced by $N$ and the resulting
state is $[N-2,1,1]$.
If two flux are removed $q$ times successively the resulting state will
be $[N-q,1,1,...,1]$, where
$1$ appears $q$ times (However, this state is not unique.
For other possibilities, see below).
The total angular momentum of this state is $3N(N-1)/2-Nq$.
The half filled state has $M_z=N(N-1)$ so
$q=(N-1)/2$. For the $\nu=2/3$ state $M_z=3N(N-1)/4$ so $q=3(N-1)/2$.
For $\nu=n/(2n\pm1)$ we find $q=\frac{(N-1)}{2}(3-\frac{1}{\nu})$.

\subsection{Trial wave functions for $N\geq 9$}
As the number of electrons increases, other choices of occupations are allowed
than proposed previously.
For $\nu=1/2$, unlike the case for $N \leq 7$, there are two 
possible compact states at $N=8$ with the correct 
value of $M_z=N(N-1)+M^*$:
The first state is $[4,2,1,1]$ with occupied CF states 
$(n,m)=(0,0),(0,1),(0,2),(0,3),(1,-1),(1,0),(2,-2),(3,-3)$, and
the second state is $[3,3,2]$ with
$(n,m)=(0,0),(0,1),(0,2),(1,-1),(1,0),(1,1),(2,-2),(2,-1)$ 
(in both cases $\sum_i m_i= M^*=0$).
Note that the second state does not have the proposed form 
$[N/2,2,1,...,1]$, but  it also turns out
its energy is almost degenerate with that of $[4,2,1,1]$
(the energies are $6.025$ vs. $6.095$).  
For $\nu=1/2$, the possible choices of $[N_0,N_1,\cdots]$ are given
in the Table \ref{energy} for several values of $N$.
Possible choices $[N_0,N_1,\cdots]$ at other principal
values of $\nu$ are given in Table \ref{choice}.
The Monte Carlo energies of these states at $\nu=1/2$
are given in Table \ref{energy}.  We choose the state
with the lowest energy as the groundstate.

\section{Monte Carlo and Exact Diagonalization Results}
Ground state properties can be calculated from the proposed trial
wave functions.  Here we concentrate on occupation numbers and
ground state energies.

\subsection{Monte Carlo calculation of occupation numbers}
For Laughlin states the edge occupation number
$n(k)\sim (k-k_{ed})^{(1/\nu-1)}$\cite{occ1,wen,occ2,occ3}.
In disk geometry single particle angular momentum $m$ is related to
the wave vector through $k=2\pi m/R_m$, where $R_m=\sqrt{2(m+1)}$.  
The value of $m$ corresponding to the location of the edge is
$m_{ed}=2N-1$\cite{yang1} for $\nu=1/2$.
The corresponding $k$ is defined as $k_{ed}$.
The exponent, $1/\nu$, in this expression 
is intimately related to the short range correlations 
of the Laughlin liquid.  Therefore one can expect to gain information about 
the the short range correlations
from  the edge occupation numbers.
This is the motivation for investigating occupation numbers.

Now we evaluate the one-particle density matrix $n(z,z^\prime)$,
which is defined as
\begin{eqnarray}
n(z,z^\prime)=N \int d^2z_2 \cdots d^2z_N
\psi^*(z,z_2,\cdots,z_N) \psi(z^\prime,z_2,\cdots,z_N) / Q_N ,
\end{eqnarray}
where the normalization integral is $Q_N = \langle \psi|\psi \rangle$.
We are especially interested in the angular dependence of the one-particle
density matrix, $n(r,re^{i\theta})$, where $r=|z|$.
Using $n(r,re^{i\theta})=\int d^2z_1 (1/2\pi |z_1|) \delta(|z_1|-r)
n(z_1,z_1 e^{i\theta})$, we have
\begin{eqnarray}
n(r,re^{i\theta}) = 
\sum_i \int d^2z_1 \cdots d^2z_N \frac{\delta(|z_i|-r)}{2\pi |z_i|}
\psi^*(z_1,\cdots,z_i,\cdots,z_N)
\psi(z_1,\cdots,z_i e^{i\theta},\cdots,z_N) / Q_N ,
\end{eqnarray}
It is convenient to perform the integration by Monte Carlo sampling with
the probability distribution
$P(z_1,z_2,\cdots,z_N)=|\psi(z_1,z_2,\cdots,z_N)|^2/
\langle \psi | \psi \rangle$, so that
\begin{eqnarray}
n(r,re^{i\theta}) = 
\sum_i \int d^2z_1 \cdots d^2z_N  \frac{\delta(|z_i|-r)}{2\pi |z_i|}
\frac {\psi(z_1,z_2,\cdots,z_i e^{i\theta},\cdots,z_N)} 
{\psi(z_1,z_2,\cdots,z_i,\cdots,z_N)} P(z_1,z_2,\cdots,z_N).
\end{eqnarray}

The angular-momentum occupation number is determined\cite{occ1} by
\begin{eqnarray}
n_m = \frac{1}{|\phi_m(r)|^2} \sum_{j=0}^{m_{eff}} \frac{e^{i\theta_j m}
n(r,re^{i\theta_j})}{m_{eff}+1},
\end{eqnarray}
where $m_{eff}$ is the highest power of $z_i$, and
\begin{eqnarray}
|\phi_m(r)|^2 = \frac{1}{2\pi m!} \frac{r^{2m}}{2^m}e^{-r^2/2}.
\end{eqnarray}

The interaction energy can also be estimated from these wave functions.
The interacting part of the the Hamiltonian is 
\begin{eqnarray}
H= \sum_{i>j} { 1 \over |z_i - z_j|}.
\end{eqnarray}

Monte Carlo sampling has been carried out through Metropolis algorithm
by changing the position of electrons,
$z_i \rightarrow z_i + \delta z_i$,
where typically $|\delta z_i| \leq 0.2-0.4$.
The wave functions vary rapidly with respect to the positions of electrons
so that after one Monte Carlo sweep
$|\psi|^2_{\rm new}/|\psi|^2_{\rm old}$
changes up to $\sim 10^{10}$ or $10^{-10}$.
It is remarkable that, in spite of this rapid variation of the wave functions,
Monte Carlo method quite nicely gives accurate  expectation values.

\subsection{Numerical Results}
\subsubsection{Wave function overlaps}

The justification of the proposed wave functions at $\nu=1/2$
comes from small-size exact diagonalization.
For $N=5$ and for the Coulomb interaction the overlap of this ground state
wave function with the exact wave function 
is $0.9989$ \cite{yang3}.
(The accuracy of the proposed trial wave function is comparable to
that of the Laughlin state: for 
four particles with Coulombic repulsions the overlap is 0.979 for the
$\nu=1/3$ Laughlin state \cite{laug}).
For $N=5$ trial wave functions of the first, second,
and third  excited states with the change in the total
angular momentum $\delta M_z=1$ can be  constructed from
the ground state wavefunction.  The overlaps with the exact ones are 
$0.9990$, $0.9663$, and $0.9098$, respectively\cite{yang3}.

\subsubsection{Occupation numbers and ground state energies}
For $N$ bigger than $5$ the proposed wave function of
Eq.(\ref{generalwaveftn}) can be tested against
the exact diagonalization results by comparing the occupation numbers.

\begin{figure}
\hskip -0.4cm
\epsfxsize=16cm \epsfysize=16cm \epsfbox{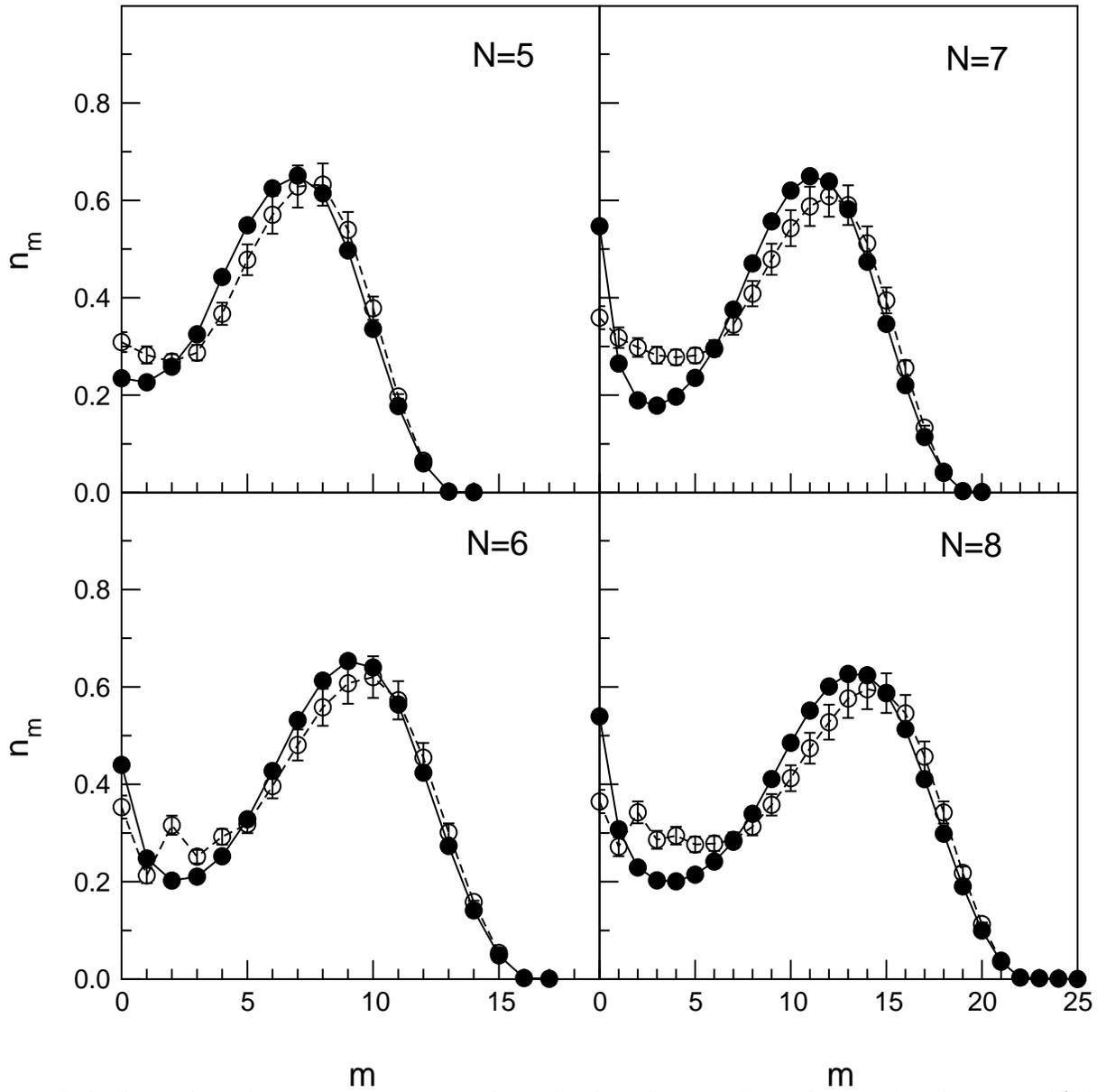}
\caption{Comparison of the result of the exact diagonalization with that of the trial wave function
at $\nu=1/3$ for $N=5,6,7$, and 8. Filled circles represent the
angular momentum occupation numbers obtained by the exact diagonalization,
while the open symbols represent these quantities evaluated by a Monte
Carlo method using the trial wave functions.}
\label{occ567L}
\end{figure}

Fig.~\ref{occ567L} displays occupation numbers for $\nu=1/3$ disk.
The results obtained from the Laughlin wave function and those obtained
from exact diagonalizations are shown.  The 
agreement between the two are excellent near the edge.
This implies that the Laughlin
wave function captures the short range correlations remarkably well.
Note that near the center of the disk a small discrepancy exists between
the results of Laughlin wave function and exact diagonalization.  

\begin{figure}
\hskip -0.4cm
\epsfxsize=16cm \epsfysize=16cm \epsfbox{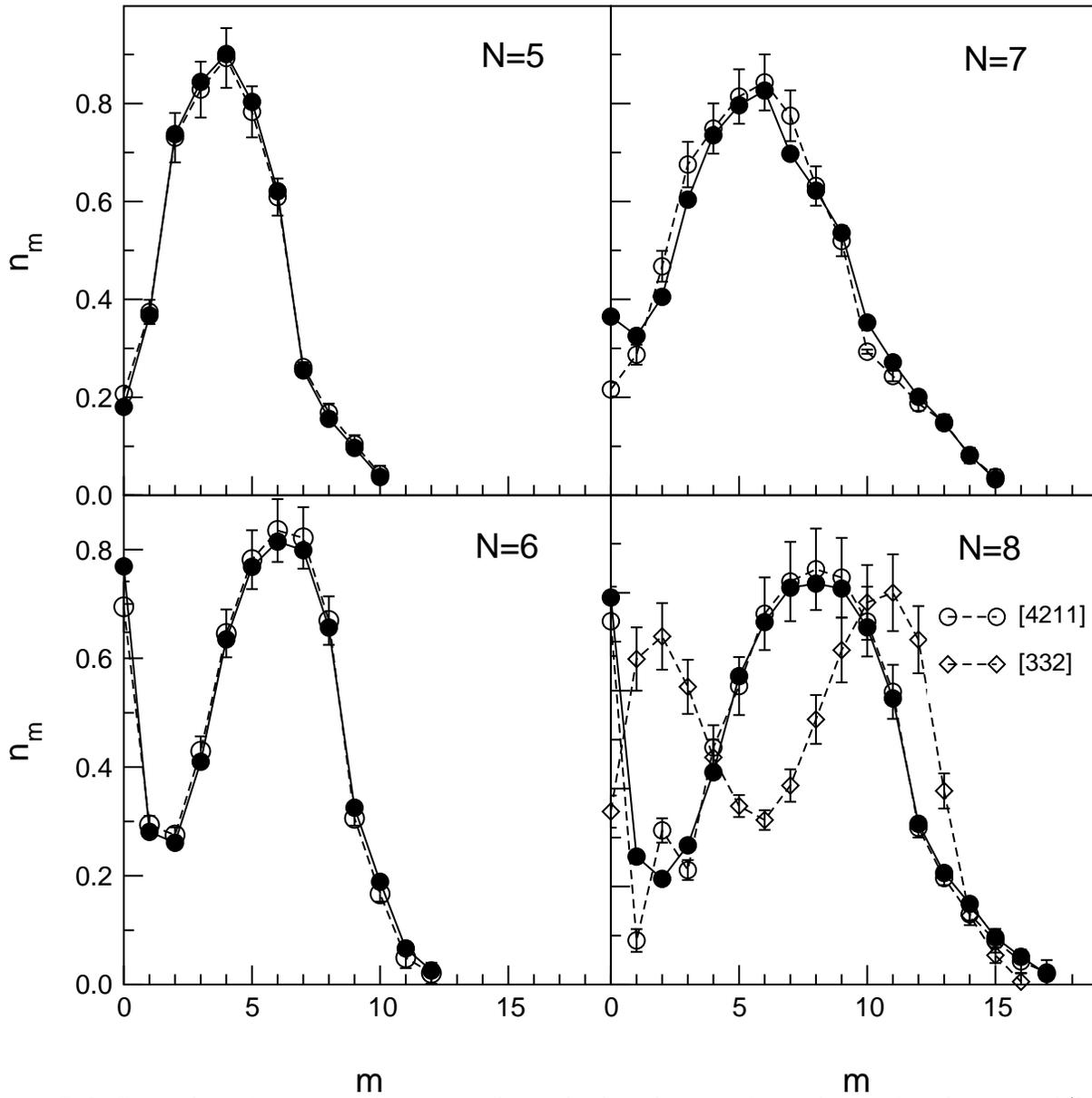}
\caption{Comparison of the result of the exact diagonalization with that of the trial wavefunction
at $\nu=1/2$ for $N=5,6,7$, and 8. Filled circles represent the
angular momentum occupation numbers obtained by the exact diagonalization,
while the open symbols represent these quantities evaluated by a Monte
Carlo method using the trial wave functions.
Note that for $N=8$ there are two possible choices of the trial
wave functions, among which [4,2,1,1] has lower Monte Carlo energy.}
\label{occ567}
\end{figure}

\begin{figure}
\hskip -0.4cm
\epsfxsize=16cm \epsfysize=16cm \epsfbox{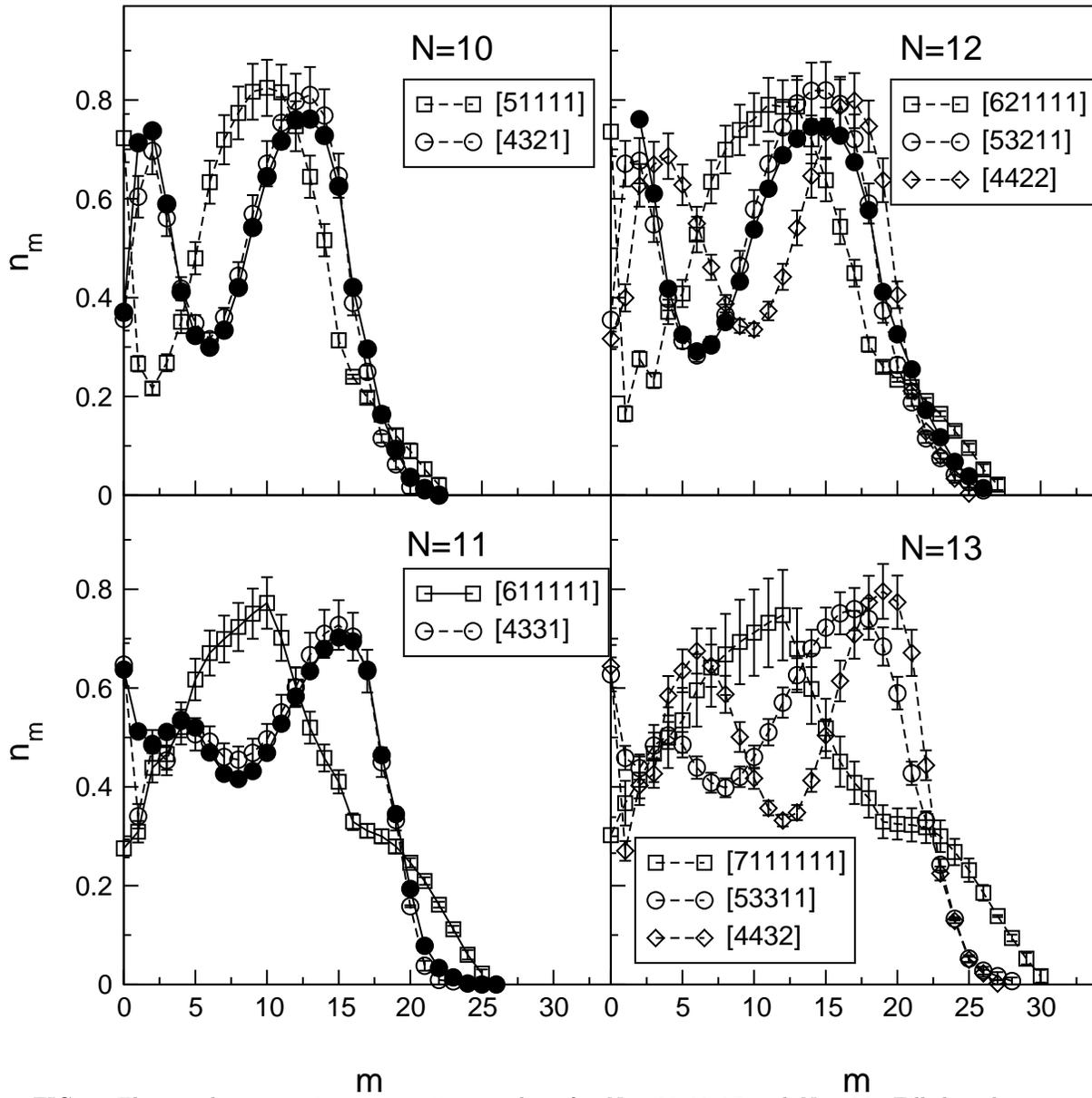}
\caption{The angular momentum occupation numbers for $N=10,11,12$ and $N=13$.
Filled circles represent the
angular momentum occupation numbers obtained by the exact diagonalization,
while the open symbols represent these quantities evaluated by a Monte
Carlo method using the trial wave functions.
Among the possible  trial wave functions, the lowest energy
states give consistent results with those of the exact states.}
\label{occ101112}
\end{figure}

Fig.~\ref{occ567} displays occupation numbers for $\nu=1/2$
disk, calculated using Eq.(\ref{generalwaveftn}).
Again, as in the case of the Laughlin wave function,
the results obtained from the 
proposed trial wave function and those of the exact diagonalization
are in excellent agreement.
For $N=8$ the state $[3,3,2]$ has the correct angular momentum
consistent with $\nu=1/2$,
but its occupation numbers show a significant deviation
from the exact ones even near the edge.
In contrast the state $[4,2,1,1]$, given by Eq.(\ref{generalwaveftn}),
agrees with the exact results.
For $N=9$ the energies for two possible wavefunctions $[5,1,1,1,1]$
and $[3,3,3]$ are nearly the same, deviating from the exact
diagonalization result by  about 1$\%$.
We find neither  $[5,1,1,1,1]$ nor $[3,3,3]$ give accurate occupation numbers.
In this case even the exact diagonalization procedure converges very
slowly.
Probably an elaborated linear combination of wave functions
could lower the energy.
For $N=10, 11, 12$, we find, among possible choices of wave functions,
that the results of the lowest energy states agree with the exact
diagonalization result (see Table \ref{energy} and  
Fig.~\ref{occ101112}).  
Note that exact diagonalization results are not available for $N \geq 13$.

\subsubsection{Zeroes of wavefunction}
\begin{figure}
\hskip -0.4cm
\epsfxsize=18cm \epsfysize=9cm \epsfbox{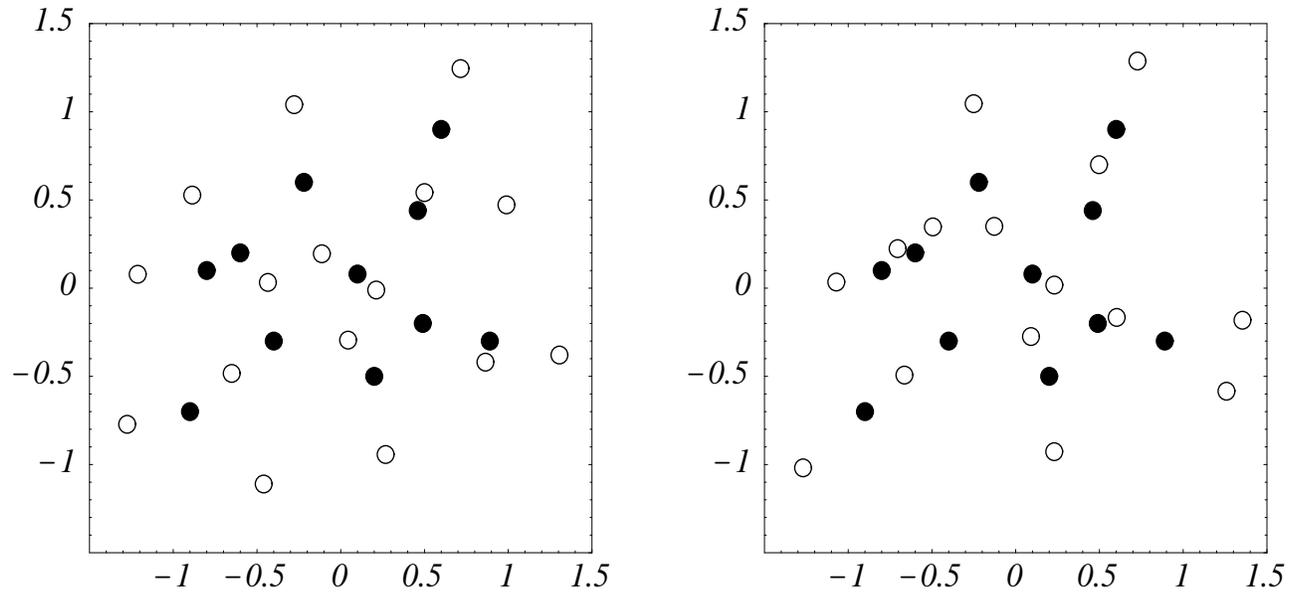}
\caption{A plot of zeros($\circ$) for a random distribution of
electrons($\bullet$) for $\nu=1/2$ and $N=12$.}
\label{zeroes}
\end{figure}

For $N=12$ the trial wavefunction $[5,3,2,1,1]$
has the lowest Monte Carlo energy.  We have investigated 
zeroes of this trial wavefunction (see Fig.\ref{zeroes}).
We find that in $[5,3,2,1,1]$ zeroes are closer to
the electrons in comparison to $[6,2,1,1,1,1]$.
This is consistent with the fact that 
$[5,3,2,1,1]$ has lower energy than $[6,2,1,1,1,1]$.
Properties of zeroes of other  trial wavefunctions are discussed
at length in Han and Yang \cite{yang2}.

\section{Field theoretical calculation of occupation numbers}
Small-size exact diagonalization calculation at $\nu=1/2$ indicates
that $n(k)$ near the edge depends linearly on
$k$ for $k \leq k_{ed}$\cite{yang1}.  We investigate using a field 
theoretical approach  whether this feature survives in the thermodynamic limit. 
The hydrodynamic action for the edge mode of a half-filled
droplet has been written down by Lee and Wen\cite{leewen},
and independently by Lopez and Fradkin\cite{frad}. 
Ignoring the coupling to the bulk degrees of freedom, the action is given by
\be
S_{edge} = {1\over 2\pi}\int \dpdx \left(\dpdt\!-\!v_c \dpdx \right)dxdt-
      {1\over 4\pi}\int \dedx \left(\dedt+v_n \dedx\right)dxdt,
\label{eq:edge_action}
\ee
where $\partial_x \phi$ and $\partial_x \eta$ are the
displacement fields of the charge mode and the neutral mode.
Following Lopez and Fradkin's approach one would have 
to add another neutral mode in the action.
The results for the occupation number remain 
unaltered, and we choose to present our result with the two-mode
model for simplicity.
The boson fields $\phi (x)$ and $\eta(x)$ satisfy
the following commutation relation
\be
[\dpdx, \phi (y)]=\pi i \delta (x-y); \,\,\,\,
[\dedx, \eta (y)]=-2\pi i \delta (x-y).
\ee
The electron operators are constructed as\cite{leewen,frad} 
\be
\psi_e (x)\sim e^{-2i\phi(x)\pm i\eta(x)}.
\label{eq:bosonized_el}
\ee
One can verify that the operators satisfy the necessary
anticommutation relations. 

The occupation number profile follows from the equal-time
single-particle Green's function, 
$G_e (x)=\langle \psi^{\dag}_e (x) \psi_e (0)\rangle$. Since the dynamics of 
$\phi$ and $\eta$ are independent, one obtains 
\be
G_e (x)\sim \exp [ G_{\phi} (x) + G_{\eta} (x) ] 
\label{eq:G_e}
\ee
where
\ba 
G_\phi (x)&=&
-2\ln \left(
{{1\!-\!\exp[-2\pi(1\!-\!ix\Lc)/\Lc L]}\over
{1\!-\!\exp[-2\pi/\Lc L]}}\right),
\nonumber \\
G_\eta (x)&=&-\ln \left(
{{1\!-\!\exp[-2\pi(1\!+\!ix\Ln)/\Ln L]}\over
{1\!-\!\exp[-2\pi/\Ln L]}}\right)
\ea
are equal-time Green's functions of $\phi$ and $\eta$.
A finite circumference $L$ of the droplet has been introduced,
as well as some high momentum 
cut-off $\Lc$/$\Ln$ for charged/neutral modes.

\begin{figure}
\hskip -0.4cm
\epsfxsize=16cm \epsfysize=16cm \epsfbox{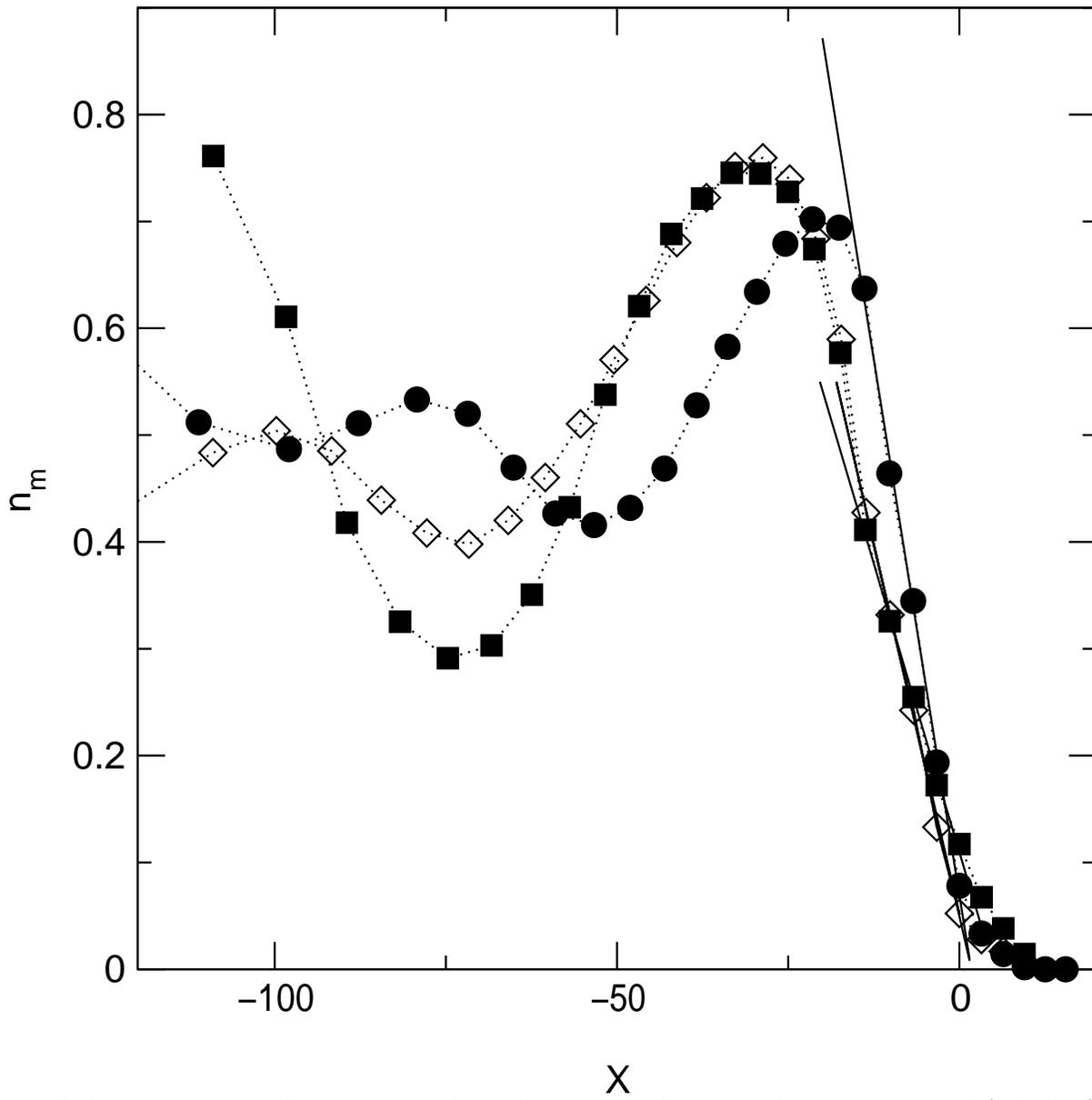}
\caption{The edge occupation numbers are fitted with those of a field
theoretical method, $Ax+B$ (solid lines).
Filled circles and squares represent the exact diagonalization results
for $N=11$ and $N=12$, respectively,
and open diamonds represent the Monte Carlo result for $N=13$ of
the trial wave function [5,3,3,1,1]}
\label{lin.occ.}
\end{figure}

A Fourier transform of 
Eq.\ (\ref{eq:G_e}) should give the momentum occupation number.
But it is first necessary to define the position of the edge.
It is natural to define the position of the edge as $m_{ed}$ from the   
the following simple model for the edge:  Consider 
a disk of electrons at half filling 
with 
uniform  occupation numbers  given by
$n(m)=1/2 (0)$ for $m\leq m_{ed}(else)$. 
The sum over occupied $n(m)$ is $N$ so  $m_{ed}=2N-1$.
A  uniform disk at $\nu=1/3$ has the occupation numbers 
given by $n(m)=1/3 (0)$  for $m\leq 3N-1(m>3N-1)$,
implying that the edge is at  $m_{ed}=3N-1$.
We have used these values of $m_{ed}$ as the positions of the edges.
In real systems effects neglected in this simple model will change
the shape of occupation number around the edge.
At half filling accurate numerical results demonstrate that
the occupation number is smeared out around $m_{ed}=2N-1$,
just like at the edge of an ordinary metal.
As Yang and Han \cite{yang1}  pointed out 
this definition of the location of the edge
gives a good scaling curve (data collapse) for the edge profile
for different values of $N$ (other definitions will fail).
In the $\nu=1/3$  incompressible (non-metallic) case,
the shape of occupation number around the edge is changed differently:
it can be deduced from the Laughlin wavefunction that 
occupation numbers beyond $m_{ed}=3N-1$ are zero \cite{occ1}.
So in the incompressible (non-metallic)
state of  $\nu=1/3$ the occupation number is zero  outside the edge. 
We find that the momentum occupation number is
\be
n(m) /n(m=0) =\left\{ \begin{array}{ll}
    e^{-2\pi m/\kappa_n}, & m>0 \\
  e^{2\pi m/\kappa_c}
\{1\!-\!m(1\!-\!e^{-2\pi(\kappa_c^{-1}\!+\!\kappa_n^{-1})})\}, & m<0.
               \end{array} \right.
\label{eq:finite_occnm}
\ee
Here  $\kappa_{c,n}$ is defined 
by $\Lambda_{c,n}L$. The boundary of a compact $\nu=1/2$ droplet coincides 
with $m=0$ ($m$ is measured from the
value corresponding to the location of the edge $m_{ed}$).
We find that the occupation number falls off linearly with $m$ for the 
interior of a compact droplet, and exponentially for the outside.
Fig.~\ref{lin.occ.} shows momentum occupation 
numbers for eleven and twelve electrons at half-filling obtained
from exact diagonalization as a function of
$x=(k-k_{ed})R_{ed}$ ($R_{ed}$ is the radius of the disk corresponding to $m_{ed}$).  
Monte Carlo results are shown for $N=13$.  These results show 
that electron occupation persists in the tail region, $x>0$. 
They also show that near $x<0$ five consecutive data points
fall, to a good approximation, on a straight line, and is well approximated by a 
linear behavior $A+Bx$ for suitable choices of $A$ and $B$, in agreement with our calculation, 
Eq.\ (\ref{eq:finite_occnm}). Based on the field-theoretical calculations presented above, 
we expect that this  ``occupation tail" will persist in the thermodynamic limit. 

\section{DISCUSSIONS}
Let us first give a short summary of our results.
The analyticity of the lowest Landau level wave functions
and the relation between filling factor and the total angular momentum
severely limits the possible forms of ground state trial wave functions.
The computed occupation numbers of these wave functions
are in excellent agreement with exact diagonalization results.
This was verified for $N \le 12$ at $\nu=1/2$.

The proposed trial wave function at $\nu=1/2$ can be tested against experiment.
Low energy excited states can be constructed by following the
prescription given in Yang and Lyue \cite{yang3}.
First the center of mass motion is constructed
by multiplying the groundstate wave function with center of mass coordinate
$Z=\sum_{i=1}^{N}x_i/N$.
Then the Gram-Schmidt process can be utilized to construct
an excited state that is  orthogonal to this state and the groundstate.
Higher excited states can be constructed repeating the Gram-Schmidt process. 
Investigating small size systems indicate that several low energy peaks 
are present in the dynamic structure
factor of a disk of electrons  \cite{yang3}.
This feature is different from that of the $\nu=1$ incompressible 
state, which has only one low energy peak.  Far-infrared radiation
of a quantum dot\cite{hk} may be used to investigate these peaks.

A naive incorporation of the 
linear dependence of the occupation numbers near
the edge into  chiral Luttinger theory  
suggests a value of tunneling exponent equal to $1$.
This value apparently agrees with experiment, but Fermi statistics will 
not be recovered in this naive approach.
We believe that the tail in the occupation number beyond
the edge must be incorporated to get the correct statistics.
As far as the value of the tunneling exponent 
is concerned it seems that the important region is the linear region.

For other principal filling factors beside $\nu=1/2$
the edge profiles and dynamic structure factors can be 
investigated along the similar line developed in this paper.
It remains a theoretical challenge to test whether
the proposed trial wave function is equivalent
to some of the field theoretical approaches developed recently.
Test of the consistency of
these different approaches will deepen our understanding
the exotic half-filled Landau level.

\acknowledgements

This work has been supported by the KOSEF under grant 981-0207-085-2,
and by Korea science and engineering foundation
through the Quantum-functional Semiconductor Center at Dongguk university.

\newpage

\begin{table}
\caption{Possible choices of $[N_0,N_1,\cdots]$ at $\nu=1/2$ and
energy evaluated.
$\langle\psi_a|H|\psi_a\rangle/\langle\psi_a|\psi_a\rangle$ are the energies
for each wave function.
$E_{ED}$ are the energies obtained by the exact diagonalization method.  Numbers in the parenthesis
represent Monte Carlo error bars.}
\label{energy}
\begin{tabular}{lllll}
$N$&$M_{tot}$&$[N_0,N_1,\cdots]$&
$\langle\psi_a|H|\psi_a\rangle/\langle\psi_a|\psi_a\rangle$ & $E_{ED}$ \\
\tableline
 3&  6&  [2,1]&0.891(3) &\\
 4&  12& [2,2]&1.686(3) &\\
 5&  20& [3,1,1]& 2.534(4) &2.537\\
 6&  30& [3,2,1]& 3.569(5) &3.567\\
 7&  42& [4,1,1,1]& 4.798(6) &4.793 \\
 8&  56& [4,2,1,1]& 6.025(6) &6.023\\
  &    & [3,3,2]  & 6.095(6) &\\
 9&  72& [5,1,1,1,1]& 7.564(6)& 7.473\\
  &    & [3,3,3]&     7.568(8)&\\
10&  90& [5,2,1,1,1]& 8.986(6)& 8.927\\
  &    & [4,3,2,1]&   8.930(6)&\\
11&  110&[6,1,1,1,1,1]& 10.763(7)& 10.58\\
  &     &[4,3,3,1]&     10.593(8)&\\
12&  132&[6,2,1,1,1,1]& 12.386(9)& 12.23\\
  &     &[5,3,2,1,1]&   12.232(9)&\\
  &     &[4,4,2,2]&     12.305(9)&\\
13&  156&[7,1,1,1,1,1,1]& 14.339(14)&\\
  &     &[5,3,3,1,1]&     14.045(12)&\\
  &     &[4,4,3,2]&       14.117(13)&\\
14&  182&[7,2,1,1,1,1,1]& 16.161(16)&\\
  &     &[6,3,2,1,1,1]&   15.950(16)&\\
  &     &[5,4,2,2,1]&     15.922(16)&\\
\end{tabular}
\end{table}

\begin{table}
\caption{Possible choices of $[N_0,N_1,\cdots]$ at general filling
factors.}
\label{choice}
\begin{tabular}{lllll}
N &$\nu=2/5$&$\nu=3/7$&$\nu=3/5$&$\nu=2/3$\\
\tableline
 3&  \ -    & \ -  & \ - & \ -\\
 4&  \ -    & [3,1]& [2,1,1] & \ -\\
 5&  [4,1]  & \ -  & \ -     & [2,1,1,1]\\
 6&  \ -    & [4,2]& [2,2,1,1] & \ -\\
 7&  \ -    & [5,1,1]& [3,1,...] & \ -\\
 8&  [6,2]  & \ -    & \ -       & [2,2,1,...]\\
 9&  [7,1,1]& [6,2,1][5,4]& [3,2,1,...][2,2,2,2,1]&[3,1,...]\\
10&  \ -    & [7,1,1,1][5,5]& [4,1,...][2,2,2,2,2] & \ -\\
11&  \ -    & \ -    & \ - & \ -\\
12&  [9,2,1]& [8,2,1,1]&[4,2,1,...]&[3,2,1,...]\\
13&  [10,1,1,1]&[9,1,...]&[5,1,...]&[4,1,...]\\
14&  \ -    & \ -        & \ -  & \ -\\
15&  \ -    &[10,2,1,1,1][9,4,2][8,6,1]&[5,2,1,...][3,3,2,2,1,...][3,2,2,2,2,2,1,1]& \ -\\
16&  [12,2,1,1][11,5]&[11,1,...][8,7,1]&[6,1,...][3,2,2,2,2,2,2,1]&[4,2,1,...][2,2,2,2,2,1,...] \\
17&  [13,1,...]& \ - & \ - &[5,1,...] \\
18&  \ -      &[12,2,1,...][11,4,2,1][10,6,2]&[6,2,1,...][4,3,2,2,1,...][3,3,2,2,2,2,1,...]& \ -\\
\end{tabular}
\end{table}

\begin{references}


\bibitem{halp}B.I. Halperin, P.A.Lee, and N. Read, Phys. Rev.  {\bf47}, 7312 (1993)
\bibitem{read1} N. Read, Semicond. Sci. Technol. {\bf 9}, 1859 (1994);
E. Rezayi and N. Read, Phys. Rev. Lett. {\bf 72}, 900 (1994). 
\bibitem{recent} R. Shankar and Ganpathy Murthy, Phys. Rev. Lett. {\bf
79}, 4437 (1997).
\bibitem{pas}V. Pasquier and F. D. M. Haldane, Nucl. Phys. B {\bf516}, 719 (1998).
\bibitem{read2}N. Read, Phys. Rev. B {\bf 58}, 16262 (1998).
\bibitem{lee} D. H. Lee, Phys. Rev. Lett. {\bf 80}, 4745 (1998).
\bibitem{halpste}B. I. Halperin and Ady Stern, Phys. Rev. Lett. {\bf 80}, 5457 (1998);
A. Stern, B. I. Halperin, F.  von Oppen, and S. H. Simon, Phys. Rev. B {\bf 59}, 12547 (1999).
\bibitem{gray}M. Grayson, D.C. Tsui, L.N. Pfeiffer, K.W. West, and A.M. Chang,
Phys. Rev. Lett. {\bf 80} 1062 (1998).
\bibitem{shy}A.V. Shytov, L.S. Levitov, and B.I. Halperin, Phys. Rev. Lett. {\bf 80}, 141 (1998).
\bibitem{leewen}D. H. Lee and X.G. Wen; cond-mat/9809160.
\bibitem{frad}A. Lopez and E. Fradkin, Phys. Rev. B {\bf 59}, 15323 (1999). 
\bibitem{yang1} S.-R. Eric Yang and J. H. Han, Phys. Rev. B {\bf 57},
R12681 (1998).
\bibitem{yang2}J. H. Han and S.-R. Eric Yang, Phys. Rev. B {\bf 58}, 
R10163 (1998).
\bibitem{yang3}S.-R. Eric Yang and W.S. Lyue,  Int. J. Mod. Phys. B, {\bf 14}, 611 (2000).
\bibitem{JK} J. K. Jain and R. K. Kamilla, Int. J. Mod. Phys. B {\bf 11}, 2621 (1997).
\bibitem{yang4}S.-R. Eric Yang, A.H. MacDonald, M. D. Johnson, Phys. Rev. Lett. {\bf 71}, 3194 (1993).
\bibitem{other}G. Dev and J.K. Jain, Phys. Rev. B {\bf 45}, 1223, (1992); J.K.
Jain and T. Kawamura, Europhys. Lett. {\bf 29}, 321 (1995);
A. Cappelli, C. Mendez, J. Simonin, and G. R. Zemba, Phys. Rev. B {\bf 58}, 16291 (1998)
\bibitem{jain}J. K. Jain, Phys. Rev. Lett. {\bf 63},  199 (1989).
\bibitem{gv} S. M. Girvin and T. Jach, Phys. Rev. B {\bf 29}, 5617 (1984).
\bibitem{laug}R. B. Laughlin, Phys. Rev. Lett. {\bf
50}, 1395 (1983).
\bibitem{occ1}Sami Mitra and A.H. MacDonald, Phys. Rev. B {\bf48}, 2005 (1993);
\bibitem{wen}X.G. Wen, Int. J. Mod. Phys. B {\bf 6} 1711 (1992).
\bibitem{occ2}E. H. Rezayi and F.D.M. Haldane,  Phys. Rev. B {\bf 50}, 17199 (1993); 
\bibitem{occ3}Unlike chiral edge, narrow Hall bar at $\nu=1/3$ has power law singularities in the  
occupation number $n(k)$ not only 
near $k=\pm 3k_F$ but also near $k=\pm k_F$.
See S.-R. Eric Yang, Sami Mitra, A.H. MacDonald, 
and M.P.A. Fisher, J. Korean Phys. Soc.{\bf 49}, S10 (1996). This result is consistent 
with the general description of Luttinger liquid by Haldane. 
\bibitem{hk}For a review see D. Heitman and J.P. Kotthaus, Physics Today (June), 56 (1993).
\end{references}
\end{document}